\documentstyle[pra,tighten,aps,epsfig,multicol]{revtex}
\begin{document}
\title{Quantum Dynamics of Three Coupled Atomic Bose-Einstein
Condensates}
\author{K. Nemoto$^\dagger$\cite{kaeemail}, C. A. Holmes$^*$, G. J.
Milburn$^\dagger$ and W. J. Munro$^\dagger$}
\address{$^\dagger$Centre for Laser Science, Department of Physics,\\
$^*$Department of Mathematics,\\
The University of Queensland,
QLD 4072 Australia}
\date{\today}
\maketitle

\begin{abstract}
The simplest model of three coupled Bose-Einstein Condensates (BEC) is
investigated using a group theoretical method.   The stationary solutions are
determined using the SU(3) group under the mean field approximation.  This
semiclassical
analysis using the system symmetries shows a transition in the dynamics of
the system
from self trapping to delocalization at a critical value for the coupling
between
the condensates.
The global dynamics are investigated by examination of the stable points
and our analysis shows the structure of the stable points depends on the
ratio of
the condensate coupling to the particle-particle interaction,  undergoes
bifurcations
as this ratio is varied.
This semiclassical model is compared to a full quantum treatment,
which also displays the dynamical transition.  The quantum case has
collapse and revival sequences superposed on the
semiclassical dynamics reflecting the underlying discreteness of the
spectrum.
Non-zero circular current states are also demonstrated as one of the higher
dimensional effects displayed in this system.
\end{abstract}

\begin{multicols}{2}

\section{Introduction}

The recent creation of neutral atom Bose-Einstein condensates (BEC)
\cite{Anderson-Science269-198-95,Bradley-PRL75-1687-95,Davies-PRL75-3969-95,Mewes-PRL77-416-96,Ensher-PRL77-4984-96}
stimulated theoretical  research aimed at understanding this new state of
matter.  Models of two coupled
BECs in a two mode approximation are considered a tractable system when total
particle number is conserved and eigenstates of the two well system are
labeled by
the particle number difference between the wells.
Two coupled BECs in a symmetric double-well potential has been
analyzed with the use of the SU(2) group
\cite{Milburn-PRA55-4318-97,Corney-PRA-printing} to show the dynamical
transition from
self-trapping in one well to delocalized oscillation through both potential
wells due to the
nonlinear particle interaction.  This model in the weak coupling
region has been further shown to demonstrate $\pi$-phase oscillations
\cite{phys_rev_a_59_raghavan}, while a semiclassical functional expression
for the
three-dimensional Josephson coupling energy have been derived
\cite{phys_rev_a_57_zapata}.
This model, however, can be considered as a
special case because of its simplicity and low-dimensionality.  Any
extension of this
model significantly increase the nonlinearity
while higher dimensional effects increase the complexity of the model
structure.

These more complex systems are of interest as the richer dynamics and model
structure allows us to treat quantum states with non-zero currents, for
instance.  In the limit of large mode number $n$, Bose-Hubbard type approaches
are useful using a mean field approximation \cite{Zoller}.
However systems with intermediate numbers of modes, $2<n<100$, are complex and
models must exploit system symmetries in order to obtain solutions.
The symmetries of these groups are the SU(n) group symmetries.
In this paper we analyze a three coupled BEC system using the operator algebra
of SU(3).  This could be realized as  three spatially separated atomic
Bose-Einstein
condensates (BECs) as illustrated in the right corner of Fig.~\ref{fig1}.
Alternatively it could describe a three condensates, occupying a single
trap and distinguished
by three internal hyperfine atomic states.  The spatially separated system
could
represent a BEC confined in a three dimensional trapping potential with three
harmonic minima in the $x-y$ plane.  Tunneling is possible between all three
minima. This symmetric triple-well system represents the simplest
two-dimensional generalization of the one dimensional double-well
\cite{Milburn-PRA55-4318-97} which allows states of non-zero angular momentum.
If the nonlinear interaction between the atoms is not too large, see
\cite{Milburn-PRA55-4318-97} (that is the total number of atoms $N$ is not
too
big), we can describe this system using a minimum of three Bose modes for the
quantum field;
\begin{equation}
\hat{\psi}(x,y,t)=\sum_{j=1}^3 [c_j(t) u_j(x,y)+c_j^\dagger(t)
u_j(x,y)^*],
\end{equation}
where the $u_j(x,y)$ are an appropriate set of orthonormal single-particle
mode
functions for this potential, and the annihilation and creation operators
$c_j,c_j^\dagger$ satisfy the equal time commutation relations
\begin{equation}
 [c_i,c_j^\dagger]=\delta_{ij}.
\end{equation}

As in the double-well case \cite{Milburn-PRA55-4318-97}, we can choose
approximate single particle mode functions which are
the localized single particle ground states for each of the three wells.
We further assume that these three lowest localized states are sufficiently
well
separated from higher energy states, and that interactions between
particles do
not change this basic property of the system configuration.  Finally we assume
that the total number of particles $N$ are conserved.  These assumptions allow
us to treat this model system in a three mode approximation.  Hence, the
many-body Hamiltonian describing atomic BECs \cite{Griffin-BEC} can be
written
in terms of the mode operators as
\begin{equation} \label{H-in-b}
\hat{H} = \omega \sum _{j=1}^3 \hat{c}_j^{\dagger}\hat{c}_j
+ \Omega \sum_{j,k=1,j \neq k}^3
\hat{c}_j^{\dagger}\hat{c}_k + \chi \sum_{j=1}^3 \hat{c}_j^{\dagger}
\hat{c}_j^{\dagger}\hat{c}_j \hat{c}_j,
\end{equation}
where $\omega$ is the mode frequency, $\chi (\leq 0)$ is the two particle
interaction strength and $\Omega (\leq 0)$ is the
tunneling frequency.  The condition, $\chi \leq 0$, corresponds to atoms
with attractive interactions.  This is the Hamiltonian of our model system in
this paper.

\section{SU(3) Group Approach}

This section shows the group theoretical treatment of the system with the
Hamiltonian specified in (\ref{H-in-b}).
In order to describe this system with SU(3) generators, we extend the
Schwinger
boson method \cite{Schwinger} and we define the eight generators of SU(3) $\{
\hat{Z}_k, \hat{Y}_k, \hat{X}_k \}$ as
\begin{equation}
\begin{array}{cc}
\left\{ \begin{array}{ll}
        \hat{X}_1 = \hat{c}_1^{\dagger}\hat{c}_1 -
\hat{c}_2^{\dagger}\hat{c}_2 \\
        \hat{X}_2 = \frac {1}{3} (\hat{c}_1^{\dagger}\hat{c}_1 +
\hat{c}_2^{\dagger}\hat{c}_2
        - 2 \hat{c}_3^{\dagger}\hat{c}_3)
        \end{array} \right.
        \begin{array}{ll}
        \hat{Y}_k = i(\hat{c}_k^{\dagger}\hat{c}_{j} -
\hat{c}_{j}^{\dagger}\hat{c}_k)\\
        \hat{Z}_k = \hat{c}_k^{\dagger}\hat{c}_{j} +
\hat{c}_{j}^{\dagger}\hat{c}_k,
        \end{array}
        \end{array}
\end{equation}
where $k=1,2,3$ and $j= (k+1) \bmod 3$. We note that the two operators
$\hat{X}_i$ commute with each other.  $\hat{X}_1$ and $\hat{X}_2$ represent
particle distributions projected on the y and x axes respectively in the right
corner of Fig.~\ref{fig1}, and
$\hat{Y}_1+\hat{Y}_2+\hat{Y}_3$ corresponds to angular momentum in this
system.
The most important relation which the
generators satisfy is the Casimir invariant of SU(3), $4 \hat{N}( \hat{N}/3
+1)$, where $\hat{N}$ is the total number operator,
$\hat{N}=\sum_{j=1}^{3}\hat{c}_{j}^{\dagger}\hat{c}_j$.
The operator algebra implies three further important identities;
\begin{eqnarray} \label{oper-ident}
\left(\frac{2\hat{N}}{3}+\hat{X}_2+\hat{X_1}\right )\left
(\frac{2\hat{N}}{3}
+\hat{X}_2-\hat{X}_1\right
) &+&\frac{4\hat{N}}{3}+2\hat{X}_2 \nonumber \\
& = & \hat{Y}_1^2+\hat{Z}_1^2 \nonumber \\
2\left (\frac{2\hat{N}}{3}+\hat{X}_2-\hat{X_1}\right )\left
(\frac{\hat{N}}{3}-\hat{X}_2\right)&+&\frac{4\hat{N}}{3}-\hat{X}_2-\hat{X}_1 \nonumber \\
& = & \hat{Y}_2^2+\hat{Z}_2^2 \\
2\left (
\frac{2\hat{N}}{3}+\hat{X}_2+\hat{X_1}\right )\left(
\frac{\hat{N}}{3}-\hat{X}_2\right)&+&\frac{4\hat{N}}{3}-\hat{X}_2+\hat{X}_1
\nonumber \\
& = & \hat{Y}_3^2+\hat{Z}_3^2. \nonumber
\end{eqnarray}
With the use of SU(3) generators, we can represent the Hamiltonian
(\ref{H-in-b}) in the form
\begin{equation}\label{newH-in-b}
\hat{H} = \Omega (\hat{Z}_1 + \hat{Z}_2 + \hat{Z}_3)
+\frac{\chi}{2} (\hat{X}_1^2 + 3 \hat{X}_2^2).
\end{equation}
Here we ignore constant terms involving the conserved total number of
particles
$\hat{N}$ which do not change the dynamics of the system.

We now specifically consider the case with $\Omega=0$ and $\chi \leq 0$.  Such
condensates are necessarily limited to a small number of atoms\cite{negative}.
For attractive forces, the ground state is three-fold degenerate and in the
occupation number representation ($|m,n,N-(m+n)\rangle=|m\rangle_1\otimes
|n\rangle_2\otimes|N-(m+n)\rangle_3$) these states are
\begin{equation}\label{ei-states}
|e_1\rangle  =  |0,0,N\rangle,\;
|e_2\rangle  =  |N,0,0\rangle,\;
|e_3\rangle  =  |0,N,0\rangle.
\end{equation}
For all these states the ground state energy is $E_0=2\chi N^2/3$.

\section{Semiclassical dynamics}

We treat the three coupled BEC model using the semiclassical mean-field
approximation.
Ignoring correlations between all operators and taking expectation values
converts the eight operator differential equations for the SU(3) generators in
the quantum system into eight differential equations for real
variables in the semiclassical system.  It is, however, very difficult
to analytically solve the
full eight dimensional equations. This leads us to consider a specific
simplified
subspace of the set of eight equations by taking the symmetric
condition, $x_1=0$.
The nature of this condition will be explained in subsection $A$ below as
to see the dynamics clearly we first need to scale the semiclassical
variables.

The expectation values of generators are
distinguished by their subscripts, while the expectation value of the total
number operator is $N$. It is convenient to scale all the semiclassical
averages
by $N$. Thus we define,
\begin{equation}
x_j  =  \frac{\langle \hat{X}_j\rangle}{N}, \; \;
y_j  =  \frac{\langle \hat{Y}_j\rangle}{N}, \; \;
z_j  =  \frac{\langle \hat{Z}_j\rangle}{N}.
\end{equation}
The equations of motion can be derived from the Heisenberg equations of
motion
of the Hamiltonian (\ref{newH-in-b}), by factoring all higher order
moments.

The three degenerate ground states for $\Omega=0$ can be associated with
particular initial conditions in the semiclassical limit. To see this we first
note that if we take matrix elements of both sides of the three operator
identities in Eq. (\ref{oper-ident}) we find that
\begin{equation} \label{orbits}
\langle e_i| \hat{Y}_j^2+\hat{Z}_j^2|e_i\rangle  =   (1-\delta_{ij})2N.
\end{equation}
Using the commutation relations for $\{\hat{Z}_i,\hat{Y}_i,\hat{X}_i\}$ and
the
corresponding uncertainty relations with respect to the ground states it is
possible to show that the ground state variances in $\{\hat{Z}_i,\hat{Y}_i\}$
scale as $N$ for $N \gg 1$. In physical terms this means that the relative
fluctuations in these variables goes to zero as $N\rightarrow\infty$, as
expected for a semiclassical limit.  This indicates that in the semiclassical
limit we may approximate
\begin{eqnarray}
\langle \hat{Z}_i^2\rangle/N^2 & \approx & \langle
\hat{Z}_i\rangle^2/N^2=z_i^2\\
\langle \hat{Y}_i^2\rangle/N^2 & \approx & \langle
\hat{Y}_i\rangle^2/N^2=y_i^2.
\end{eqnarray}

The resulting semiclassical equations will be analyzed from two perspectives.
We first show a dynamical transition between self-trapping and
delocalization when initial
conditions are given by Eq. (\ref{ei-states}).  This analysis determines the
critical point for the dynamics transition.
Secondly we consider the dependence of the stable points on the ratio of the
coupling between condensates to the particle-particle interaction strength,
and
show bifurcations in the set of the stable points.

\subsection{Dynamics transition}
Using these relations in Eq. (\ref{orbits}) we can
construct semiclassical correspondences for each of the ground
states. This is shown below
\begin{eqnarray*}
\begin{array}{ccc}
|e_1\rangle     & |e_2\rangle              & |e_3\rangle\\
y_1^2+z_1^2=0\ \ \      & y_2^2+z_2^2=0\ \ \           &  y_3^2+z_3^3=0\\
x_1=0\ \ \ \    & x_1=1\ \ \ \ \            &     x_1=-1\\
x_2=-2/3\ \ \ \ & x_2=1/3\ \ \ \ \             &     x_2=1/3.\\
\end{array}
\end{eqnarray*}
It is easy to verify that these curves are invariant under the semiclassical
dynamics with $\Omega=0$. As each of the ground states is equivalent, up to a
rotation in the phase space, we will now restrict the discussion to
$|e_1\rangle$ (a condensate localized initially in the third well) without
loss
of generality, and examine the dynamics when $\Omega\neq 0$.

\begin{figure}
\center{ \epsfig{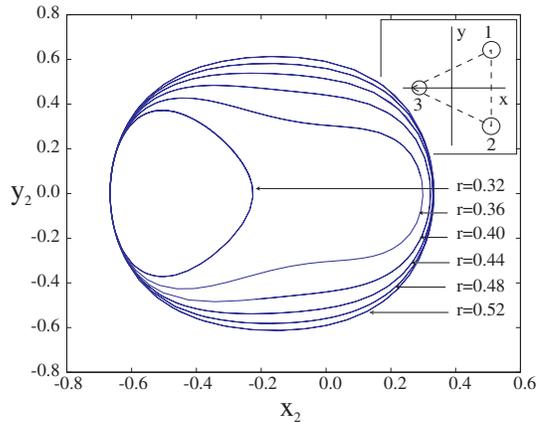}}
\caption{Phase space orbits of the semiclassical dynamics projected on the
$y_2-x_2$ plane,  for various values of $r$.  The total number of
atoms $N$ in the system was $5\times 10^{7}$. $r_*=1/3$ is the critical
value for localization. The transition from localized to delocalized
dynamics is apparent.   The sub-figure in the right upper corner is a
schematic representation
of three spatially distinct Bose-Einstein condensates located at the minima
of a potential with
triangle symmetry.  The tunneling coupling constant between all the wells
is equal and
nonzero.}
\label{fig1}
\end{figure}

Use of the initial state $|e_1 \rangle$ naturally restricts the
dynamics due to system symmetries, allowing us to study an analytically
solvable subsystem.
The initial state $|e_1 \rangle$ and the Hamiltonian (\ref{newH-in-b}) are
symmetric to
permutations of wells 1 and 2, then the resulting dynamics also satisfy
this symmetry.
This ensures that $x_1=0$ for all time in the semiclassical limit, which we
previously
referred to as the symmetric condition.
This symmetric condition specifies the subspace in which the reduced system
dynamics lies,
giving the additional conditions, $y_1=0$, $y_2=-y_3$, and $z_2=z_3$.
The semiclassical dynamics is governed by the following four dimensional
system
\begin{eqnarray} \label{semi eq}
\dot{x}_2 & = & -2\Omega y_2\nonumber \\
\dot{y}_2 & = & \Omega(3x_2+z_1-z_2)-3\chi N z_2 x_2\\
\dot{z}_1 & = & -2\Omega y_2\nonumber \\
\dot{z}_2 & = & \Omega y_2+3\chi N x_2 y_2.\nonumber
\end{eqnarray}
The initial state $|e_1 \rangle$ gives the semiclassical system the initial conditions
\begin{eqnarray} \label{initial}
x_2(0) & = & -2/3\\
y_i(0) & = & z_i(0)=0. \nonumber
\end{eqnarray}
The semiclassical dynamics governed by the equations of motion (\ref{semi eq}) is numerically
shown in Fig.~\ref{fig1}.
Fig.~\ref{fig1} shows a transition from localization to global oscillation, and this dynamics
change can be explained by the stable point analysis.
This reduced system is integrable as these equations satisfy two constants
of motion;
\begin{eqnarray}
 \Omega(z_1+2 z_2)+\frac{3\chi N x_2^2}{2} & = & H/N \\
3x_2^2+2(y_2^2+z_2^2)+z_1^2 & = & 4/3,
\label{number_cons}
\end{eqnarray}
which correspond to energy and total number conservation respectively.
The latter constant (\ref{number_cons}) follows from two stricter constraints
\begin{eqnarray}
\left( \frac{2}{3}+x_2\right)^2 &=& z_1^2\\
\label{cons1}
2\left( x_2+\frac{1}{6}\right)^2+y_2^2+z_2^2 &=& \frac{1}{2},
\label{cons2}
\end{eqnarray}
which are derived from Eq. (\ref{oper-ident}) with the symmetric condition
in the semiclassical limit.

The equations of motion with the chosen initial conditions may be solved
explicitly as a function of $x_2$ and depend on only one parameter; the
ratio of the
coupling constant $\Omega$ to the interaction strength $\chi$,
\begin{equation}
r=\frac{\Omega}{N\chi} \; (\geq 0).
\end{equation}
The solutions are,
\begin{eqnarray} \label{solutions}
z_1(t) & = & x_2(t)+\frac{2}{3} \nonumber\\
z_2(t) & =  &
-\frac{x_2(t)}{2}-\frac{3x_2(t)^2}{4r}+\left(-\frac{1}{3} +
\frac{1}{3r}\right)\\
y_2(t)^2 & = & f(x_2(t)), \nonumber
\end{eqnarray}
where
\begin{eqnarray}
f(x_2(t))  &=&  -\left (a(r) -\left (x_2(t)+\frac{r}{3}\right
)^2\frac{3}{4r}\right )^2 \nonumber \\
&\;&\;\;\;\;+\frac{2r}{3}\left
(x_2(t)+\frac{r}{3}\right )(2-\frac{1}{r})+B(r),
\end{eqnarray}
and where the integration constants are given by
\begin{eqnarray}
a(r) & = & -\frac{4r}{3}+\frac{(r-2)^2}{12r}\\
B(r) & = & \left (\frac{4r}{3}\right )^2
-\frac{2}{9}(2r-1)(r-2).
\end{eqnarray}

The solutions are oscillatory and only exist for $-2/3\leq x_2 \leq 1/3$,
and are strongly dependent on the roots of the function $f(x)$.
$f(x_2)$ is fourth
order in terms of
$x_2$ and can have up to four real roots.  For increasing $r$, the root
structure
of $f(x_2)$ changes.  To really understand these changes it is
instructive to look at the fixed point structure of system (\ref{semi
eq}). If
all the derivatives of the equations of motion (\ref{semi eq})
are zero then,
\begin{eqnarray} \label{s_eq1}
& &y_2=0,\nonumber\\
& &-3x_2-z_1+z_2+\frac{3}{r}z_2x_2 = 0.
\end{eqnarray}
Taking the positive root for $z_1$ in the constraint (\ref{cons1}),
$z_1=x_2+2/3$,
the second equation can be written
\begin{equation} \label{s eq}
-4x_2+z_2+\frac{3}{r}z_2x_2-\frac{2}{3}=0.
\end{equation}
So the fixed points are the crossing points of this
equation
with the constraint,
\begin{equation} \label{cons}
2\left( x_2+\frac{1}{6}\right)^2+z_2^2=\frac{1}{2},
\end{equation}
obtained by applying $y_2=0$ to constraint (\ref{cons2}).
Together these result in the following fourth order ploynomial in $z_2$
\begin{eqnarray} \label{C}
\frac{18}{r^2}\left( z_2-\frac{2}{3}\right) \left(
z_2^3+\frac{2}{3}(1-4r)z_2^2-2r(1-r)z_2+\frac{4r^2}{3}\right) =0.
\end{eqnarray}

\begin{figure}
\center{ \epsfig{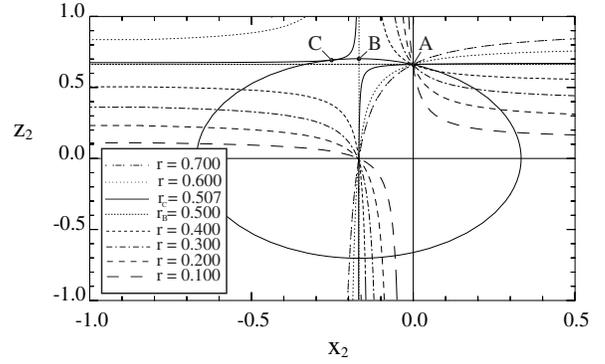}}
\caption{The $r$ dependence of the stable points.  Each
hyperbolic curve shows equation (\ref{s eq}) for different $r$, while the ellipse
represents constraint (\ref{cons}).  Point $A$ is a stable point for any $r$.  Point $B$
shows when the equation (\ref{s eq}) collapses into the two lines, and
point $C$ is the unique point of tangential contact to the ellipse.}
\label{fig_a1}
\end{figure}

Fig.~\ref{fig_a1} shows the curves (\ref{s eq}) and (\ref{cons})
in the $x_2$--$z_2$ plane for various values of $r$.
The first factor of (\ref{C}) gives us the point $A$
($x=0,z_2=\frac{2}{3}$).
For larger $r$ only one branch of the hyperbola (\ref{s eq}) intersects the
constraint and there
are just two elliptic fixed points, one at A and one in the third
quadrant.  The solutions
with initial condition $x_2=-\frac{2}{3}$ are far away from the elliptic
fixed points and perform large delocalised oscillations on the sphere
(\ref{cons2}).  A projection of one such solution is shown in Fig.~\ref{fig1} for
$r=0.52$.
As $r$ is decreased a second branch
touches the constraint at C Fig.~\ref{fig_a1}.
This occurs when the second factor in  (\ref{C}) has a double root at $r=0.507425$.
Two new fixed points, a saddle and a center, then appear and
move apart as $r$ is decreased away from $0.507425$. The stable and unstable
manifolds of the saddle intersect, forming a lopsided figure eight like
separatrix with the new elliptic fixed point in the one lobe and the elliptic
fixed point at A in the other. When $r$ is $\frac{1}{2}$ the hyperbola
reduces to the lines $z_2=\frac{2}{3}, x_2=- \frac{1}{6} $ and the
solutions in $(x_2,y_2)$ space are symmetrical about $x_2=-\frac{1}{6}$.
If $r$ is further decreased the separatrix approaches our initial
condition (\ref{initial}). At $r=\frac{1}{3}$ the initial
condition actually
lies on the
separatrix. This occurs when the
unstable fixed point lies on the solution curve (\ref{solutions}),
which amounts to looking for a double root of $f(x_2)$. If we
let $u=x_2+\frac{r}{3}$, then
\begin{equation}
f(u-\frac{r}{3})=-\left (a(r)-\frac{3 u^2}{4r}\right )^2+\frac{2u}{3}(2r-1)+B(r).
\end{equation}
and we require $f=0$ and $f^\prime=0$.
$f^\prime=0$ gives,
\begin{equation}
u^2=\frac{4r}{9}\left (a(r)-\sqrt{4a(r)^2-3B(r)}\right ),
\end{equation}
which when substituted into $f=0$ gives only one real solution at
$r_*=1/3$, $x_2=0$.
Now for $r>\frac{1}{3}$ the solutions
lie within the separatrix and so
as $r$ is decreased through $\frac{1}{3}$ the oscillations suddenly
reduce to half their former size as can be clearly seen in
Fig.~\ref{fig1}.

In the physical space of the potential then, the condensate
remains localized at the bottom of the first well,$x_2=-\frac{2}{3}$, for $r=0$ .  As we increase $r$ from
zero, $x_2(t)$ begins to oscillate. In the $x_2-y_2$ plane,
the point is replaced by small oscillations, within the separatrix,
which pass through
$x_2=-2/3$. This is referred to as dynamical localization.  Even though the
coupling between wells is present, the nonlinear interaction between particles
prevents the condensate from moving away from its initially localized state.
Eventually for the critical value of $r_=\frac{1}{3}$ the orbit in the $x_2-y_2$ plane
extends across both half planes for positive and negative $x_2$
values and the condensate is no longer localized.

\begin{figure}
\center{ \epsfig{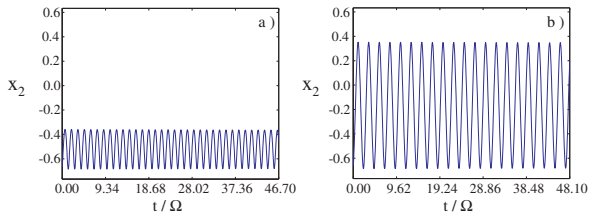}}
\caption{Time evolution of $x_2(t)$ in the semiclassical
regime versus time below (Figure a) and above (Figure b) the critical
value $r_*=1/3$ for localization. Figure a) corresponds to the below
threshold regime with
$r=0.283$ while figure b) corresponds to the above threshold regime with
$r=0.506$.  The number of atoms is fixed at $N=50$, and the time $t$ is
normalized
by $\Omega$.}
\label{fig2}
\end{figure}

In Fig.~\ref{fig1}  we show the phase space orbits of the semiclassical
dynamics
projected onto the $x_2-y_2$ plane for various values of the parameter $r$.
The
initial conditions correspond to a condensate localized in well 3. The
transition from localized to delocalized motion is seen when the orbit in the
phase space extends into the $x_2>0$ half of the plane. In Fig.~\ref{fig2} we plot
$x_2(t)$
as a function of time for the condensate initially localized in well
3 with the same initial conditions as in Fig.~\ref{fig1} for two
values of $r$, one above and below the critical value $r_*$ for localization.
The transition from localized to delocalized oscillations is apparent.

\section{Quantum dynamics}
In this section we treat the three coupled BEC model fully quantum
mechanically and numerically calculate the time evolution of the
particle distribution with the initial condition $|e_1\rangle$.
For fixed total atom number $\hat{N}$, a suitable basis of the system Hilbert
space is the number eigenstates $|n,m,N-(n+m)\rangle$ which are simultaneous
eigenstates of the generators, $\hat{X}_1$ and $\hat{X}_2$ of the Abelian
sub-algebra of SU(3).  The unitary evolution operator is given by
$U(t)=\exp \left[-i H t\right]$ where $H$ is specified by
(\ref{newH-in-b}).  The unitary evolution matrix is then computed and
the initial state $|\psi(0)\rangle=|e_1\rangle$ is then propagated forward
in time.
At each time step we compute the averages of $\langle\hat{X}_1\rangle$ and
$\langle\hat{X}_2\rangle$.   These averages show the particle distribution
projected on
the y and x axes in Fig.~\ref{fig1} respectively.
However because of the initial conditions for the state $|\psi(0)\rangle$
the average of $\langle\hat{X}_1\rangle$ does
not change and in fact remains zero.  It is not considered further here.

\begin{figure}
\center{ \epsfig{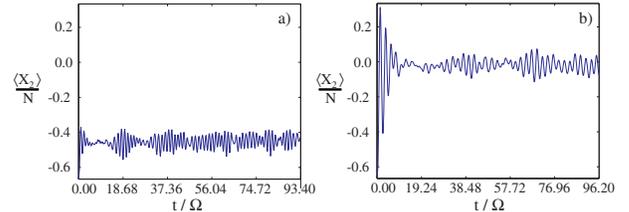}}
\caption{The quantum dynamics of $\langle \hat{X}_2\rangle$
versus time below (Figure a) and above (Figure b) the critical value
$r_*=1/3$
for localization. Figure a) corresponds to the below threshold regime with
$r=0.283$ while figure b) corresponds to the above threshold regime with
$r=0.506$.  The number of atoms is fixed at $N=50$, and the time is
normalized by $\Omega$.}
\label{fig3}
\end{figure}

Fig.~\ref{fig3} shows the evolution of a
condensate initially localized in state $|e_1\rangle$ with number of atoms
fixed
at $N=50$.  For short times we see the same oscillations as in the
corresponding
semiclassical case, both below and above the critical value $r_*$.
The oscillations of the quantum mean values decay due to the intrinsic quantum
fluctuations in the number of atoms in each individual well, while the
total particle number
in the system remains fixed.  The collapses and revivals of the
oscillations in the quantum
system arise from the discrete nature of the eigenvalue spectrum for finite
atom number.  Such
phenomena were also observed in two coupled condensates
\cite{Milburn-PRA55-4318-97}.

\begin{figure}
\center{ \epsfig{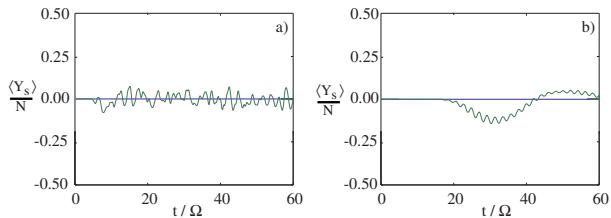}}
\caption{The quantum dynamics of $\langle \hat{Y}_s\rangle/N$
versus time with the initial state $|g_2\rangle$, where
$\hat{Y}_s = \hat{Y}_1+\hat{Y}_2 +\hat{Y}_3$ is the
angular momentum.
The angular momentum for small $r=0.5$ (Figure a) fluctuates, while
non-zero circular motion
appears in Figure b) for larger $r=1.3$.  The number of atoms is fixed at
$N=50$,
and the time is normalized by $\Omega$.}
\label{fig_a2}
\end{figure}

One of the interesting higher dimensional features of this system is the
existence of
non-zero circular current states.   For $r=0$, the
system has the three-fold degenerate ground states, $|e_1\rangle$, $|e_2\rangle$,
and $|e_3\rangle$,
and superpositions of those states create another set of orthonormal ground
states given as
\begin{eqnarray}
|g_1\rangle &=& \frac{1}{\sqrt{3}}\left( |e_1\rangle  +|e_2\rangle
+|e_3\rangle  \right)
\nonumber \\
|g_2\rangle &=& \frac{1}{\sqrt{3}}\left( e^{-i2\pi/3}|e_1\rangle
+|e_2\rangle
+e^{i2\pi/3}|e_3\rangle  \right) \\
|g_3\rangle &=& \frac{1}{\sqrt{3}}\left( e^{-i2\pi/3}|e_1\rangle
+e^{i2\pi/3}|e_2\rangle
+|e_3\rangle  \right). \nonumber
\end{eqnarray}
These states are invariant under $2\pi/3$ rotations due to system
symmetries, as
discussed in \cite{isqm}.
Taking the state $|g_2\rangle$, we examine the quantum dynamics of the
angular momentum
$\hat{Y}_s=\hat{Y}_1+\hat{Y}_2+\hat{Y}_3$.
(The state $|g_3\rangle$ gives the same dynamics, however since these
states are
anti-symmetric to each other, clockwise motion in $|g_2\rangle$ corresponds
anti-clockwise in $|g_3\rangle$.)
For $r=0$, the state $|g_2\rangle$ is the ground state, and the average
angular momentum
remains zero, though for non-zero $r$, a non-zero average angular momentum
appears as seen in
Fig.~\ref{fig_a2}.  This shows quantum dynamics of the average angular
momentum
normalized by the total number $N$ for two different values of $r$.
For small $r$ the non-zero angular momentum does not develop into any
stable circular motion
(Fig.~\ref{fig_a2}a), while circular motion can be established for large $r$
(Fig.~\ref{fig_a2}b), and becomes increasingly stable for larger $r$.

\section{Discussion and Conclusion}
In this paper we have shown how the generators of SU(3) can be used to
describe
the quantum and semiclassical dynamics of three symmetrically coupled atomic
Bose-Einstein condensates. By taking expectation values of the Heisenberg
equations of motion and factoring all higher order moments we can derive the
semiclassical mean field equations.
The nonlinear terms arising from hard collisions lead to a
dynamical bifurcation in the semiclassical dynamics as the tunneling
strength is
increased, reflecting a transition from localized dynamics to tunneling
currents.  The $r$ dependence of the system is much more complicated than
that found for
two coupled BECs.  The $r$ dependence of the stationary solutions in this
paper
constitutes only a small part of the total complexity of this system and
only some of the
higher dimensional effects.  Our quantum treatment verified the dynamical
transition found in
the semiclassical analysis.  This three coupled BEC model is the simplest
model to
have non-zero circular current states, and we have shown non-zero circular
motion can appear
given appropriate initial conditions.

The analysis of this paper focused on comparing the semiclassical dynamics
with the
evolution of quantum mean values.  This restricted analysis naturally
suggests further
examination of the dynamics of full quantum states.  However in order to
examine full quantum
states, it is necessary to use more powerful group theoretic tools and this
will be the
subject of a future paper.

\acknowledgments{KN would like to thank Australian International Education
Foundation
(AIEF) for financial support.  WJM acknowledges the support of the
Australian Research Council.}

\end{multicols}
\end{document}